\documentclass[runningheads]{llncs}
\usepackage{graphicx}
\usepackage{amsmath}
\usepackage{multicol}
\DeclareMathOperator*{\argmin}{argmin}
\usepackage{multirow}
\usepackage{diagbox}
\usepackage[table]{xcolor}

\begin{document}
\title{A hybrid deep learning framework for integrated segmentation and registration: evaluation on longitudinal white matter tract changes}
\titlerunning{A hybrid network for integrated segmentation and registration}
%
\author{Bo Li\inst{1,2}\and
Wiro J. Niessen\inst{2,3}\and
Stefan Klein\inst{2}\and
Marius de Groot\inst{2} \and
M. Arfan Ikram\inst{2}\and
Meike W. Vernooij\inst{2} \and
Esther E. Bron\inst{2}}


\authorrunning{B.Li et al.}

\institute{Northeastern University, Shenyang, China \and Erasmus MC, Rotterdam, the Netherlands \and
Delft University of Technology, Delft, the Netherlands\\
\email{b.li@erasmusmc.nl}}

\maketitle             

\begin{abstract}
To accurately analyze changes of anatomical structures in longitudinal imaging studies, consistent segmentation across multiple time-points is required. Existing solutions often involve independent registration and segmentation components. Registration between time-points is used either as a prior for segmentation in a subsequent time point or to perform segmentation in a common space. In this work, we propose a novel hybrid convolutional neural network (CNN) that integrates segmentation and registration into a single procedure. We hypothesize that the joint optimization leads to increased performance on both tasks. The hybrid CNN is trained by minimizing an integrated loss function composed of four different terms, measuring segmentation accuracy, similarity between registered images, deformation field smoothness, and segmentation consistency. We applied this method to the segmentation of white matter tracts, describing functionally grouped axonal fibers, using N=8045 longitudinal brain MRI data of 3249 individuals. The proposed method was compared with two multistage pipelines using two existing segmentation methods combined with a conventional deformable registration algorithm. In addition, we assessed the added value of the joint optimization for segmentation and registration separately. The hybrid CNN yielded significantly higher accuracy, consistency and reproducibility of segmentation than the multistage pipelines, and was orders of magnitude faster. Therefore, we expect it can serve as a novel tool to support clinical and epidemiological analyses on understanding microstructural brain changes over time.

\end{abstract}
\section{Introduction}

In longitudinal imaging studies, the consistency of segmentations can be improved by using methods tailored to longitudinal data \cite{yendiki2016joint}. Existing solutions often involve independent registration and segmentation components, which are performed sequentially or iteratively in a multi-stage pipeline. Spatial correspondence established with deformable registration is used either to introduce a prior for segmentation in a subsequent time-point, or to perform segmentation in a common space. We here propose a novel hybrid convolutional neural network (CNN) that optimizes segmentation and registration in a single procedure. We hypothesize that the joint optimization leads to increased performance on both tasks.

This work is one of the first learning-based frameworks for joint optimization of segmentation and registration. Existing methods for joint optimization, e.g., using a Markov Random Field \cite{parisot2014concurrent} or Expectation Maximization \cite{pohl2005expectation} framework, rely on non-learning based registrations, and therefore need to be optimized on test data. In addition, there are two types of work that are closely related but are different from joint optimization. The first type of methods focus on registration-based segmentation \cite{yezzi2003variational}, e.g., atlas-based segmentation and contour propagation. These methods label the images by registering atlas images to the data to be segmented. An example of segmentation-based registration is \cite{hu2018label}. The second type of methods aim at improving segmentation with predefined registration. For instance, deep learning-based segmentation methods apply pre-estimated transformations to introduce labels for a weakly supervised task \cite{vlontzos2018deep}. An example of improving registration with pre-segmented images can be found in \cite{balakrishnan2019voxelmorph}. In contrast to the above methods, our hybrid method does not require online-optimization or any prepared label and transformation. Segmentation and registration are performed at the same time resulting in a segmented structure, a transformation between images, and a deformed image.

We propose the hybrid CNN in a general and cross-sectional manner in Section~\ref{s:2}. The method is demonstrated on the segmentation of white matter tracts, describing functionally grouped axonal fibers, using a large diffusion-weighted MRI (DWI) dataset. We evaluate its performance in a longitudinal setting with multiple time-points per individual. In Section~\ref{s:3}, we compare the hybrid method with two multistage pipelines, and assess the added value of joint optimization for segmentation and registration separately. A discussion of our method and the results can be found in Section~\ref{s:4}.

\section{Hybrid method}
\label{s:2}
Let $I(x)$ be an input image, $S(x)$ be its segmentation label and $x \in R^3$ denote the spatial coordinates. In structure segmentation, parameters $\boldsymbol{\Theta}$ for a function $\mathcal{F}_{\boldsymbol{\Theta}}$ are estimated such that 
\begin{equation}\label{eq1}
 S = \mathcal{F}_{\boldsymbol{\Theta}}\big(I\big).
\end{equation} 
For registration of images, a transformation $\boldsymbol{\mathcal{T}}$ is applied to an image (source image, $I_s$) to optimally fit another image (target image, $I_t$), i.e., for $\forall x \in R^3$, $I_t(x)$ and $I_s\big(\boldsymbol{\mathcal{T}}(x)\big)$ correspond to a same anatomical location:
\begin{equation}\label{eq2}
I_t(x) \approx I_s\big(\boldsymbol{\mathcal{T}}(x)\big).
\end{equation} 

The transformation can be written as $\boldsymbol{\mathcal{T}}_{t,s}(x)$, or in short as $\boldsymbol{\mathcal{T}}(x)$. $\boldsymbol{\mathcal{T}}(x)$ includes both global (rigid, affine) and local (non-rigid) transformations. For any pair of inputs, $\boldsymbol{\mathcal{T}}(x)$ is estimated by a shared function $\mathcal{G}_{\boldsymbol{\Phi}}$, i.e., $\boldsymbol{\mathcal{T}}_{t,s}(x) = \mathcal{G}_{\boldsymbol{\Phi}}(I_t, I_s)$. Parameters $\boldsymbol{\Phi}$ for $\mathcal{G}_{\boldsymbol{\Phi}}$ are globally optimized on all training data.

\begin{figure}[!]
\centering
\includegraphics[width=\textwidth]{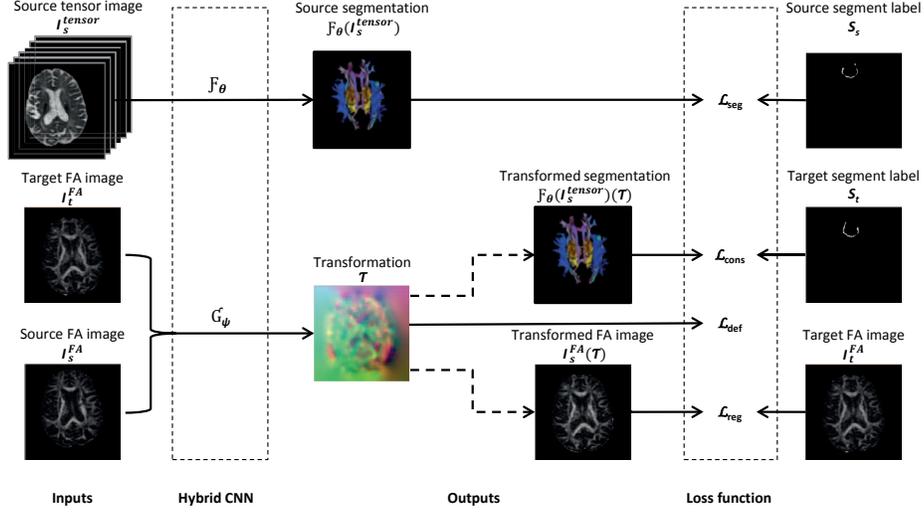}
\setlength{\belowcaptionskip}{-15pt}
\caption{Overview of the method. $\boldsymbol{\Theta}$ and $\boldsymbol{\Phi}$ denote the parameters of the segmentation and registration task, respectively. Both the source tensor and FA images were estimated from the same diffusion-weighted MRI scan. The loss function consists of $\mathcal{L}_{seg}$, $\mathcal{L}_{cons}$, $\mathcal{L}_{def}$ and $\mathcal{L}_{reg}$ terms.}\label{fig1}
\end{figure}

In this work, we integrate the parameters $\boldsymbol{\Theta}$ and $\boldsymbol{\Phi}$ in a single hybrid CNN. The overview of our method is illustrated in Fig.~\ref{fig1}. We describe the loss function and network architecture in following paragraphs.

\subsubsection{Loss function.}
\label{sss:2.1}

The loss function for optimization of $\boldsymbol{\Theta}$ and $\boldsymbol{\Phi}$ is composed of four terms that measure segmentation accuracy ($\mathcal{L}_{seg}$), similarity between registered images ($\mathcal{L}_{reg}$), deformation field smoothness ($\mathcal{L}_{def}$) and segmentation consistency ($\mathcal{L}_{cons}$), respectively, i.e.,
\begin{equation}\label{eq3}
    \begin{gathered} 
     \boldsymbol{\hat{\Phi}},\boldsymbol{\hat{\Theta}} = \argmin_{\boldsymbol{\Phi}, \boldsymbol{\Theta}} \{\mathcal{L}_{seg} \big(S_s, \mathcal{F}_{\boldsymbol{\Theta}}(I_s)\big)+\alpha \mathcal{L}_{reg} \big(I_t, I_s(\boldsymbol{\mathcal{T}})\big)+\beta \mathcal{L}_{def}(\boldsymbol{\mathcal{T}}) \\
     +\gamma \mathcal{L}_{cons} \bigg(S_t, \mathcal{F}_{\boldsymbol{\Theta}}(I_s)\big(\boldsymbol{\mathcal{T}}\big)\bigg)\},
    \end{gathered}
\end{equation}
where $\boldsymbol{\mathcal{T}}$ depends on $\boldsymbol{\Phi}$. Segmentation accuracy was quantified by the agreement between the predicted segmentation of source images ($\mathcal{F}_{\boldsymbol{\Theta}}(I_s)$) and the segment labels ($S_s$). Consistency was quantified by the correspondence between the transformed segmentation of source images ($\mathcal{F}_{\boldsymbol{\Theta}}(I_s)\big(\boldsymbol{\mathcal{T}}\big)$) and the segment labels in target space ($S_t$). The segmentation accuracy and consistency terms were computed using weighted inner product metric  \cite{li2018reproducible}. Similarity between the registered images was quantified using mean squared error. Smoothness of the deformation field was encouraged by minimizing a diffusion regularization term on the estimated transformations, which defined as a mean of squares of first derivatives. We set the hyperparameters to $\alpha=10,\beta=0.1$ and $\gamma=1$.

\subsubsection{Network architecture.}
\label{ss:2.2}
The hybrid 3D CNN models $\mathcal{F}_{\boldsymbol{\Theta}}$ and $\mathcal{G}_{\boldsymbol{\Phi}}$ in parallel by a series of convolutions and non-linearity operations, using similar U-Net architectures \cite{ronneberger2015u} with skip connections. The encoder paths were gradual compression processes of extracting abstract features of the structure for $\mathcal{F}_{\boldsymbol{\Theta}}$, and of estimating global deformations between images for $\mathcal{G}_{\boldsymbol{\Phi}}$. The decoder paths restored the details in segmentation ($\mathcal{F}_{\boldsymbol{\Theta}}$) and refined local deformations ($\mathcal{G}_{\boldsymbol{\Phi}}$) by decompressing features and combining them with the shallow information at the same scales. The convolution layers produced a set of $k$ feature maps by individually convolving the input with $k$ kernels. In this work, we used $k=[16,32,64,128,256,128,64,32,16]$ for both $\mathcal{F}_{\boldsymbol{\Theta}}$ and $\mathcal{G}_{\boldsymbol{\Phi}}$. Each convolution layer was of kernel size $(3,3,3)$, and followed by a batch normalization and a leaky ReLu layer ($a = 0.2$) for non-linerities. Resamplings were performed using max-pooling and up-sampling operations.

The last layer of $\mathcal{F}_{\boldsymbol{\Theta}}$ was an softmax function resulting in a posterior probability $P\big(S(x)|\boldsymbol{\Theta},I(x)\big)$. During performance evaluation, the probabilistic map was binarized with a threshold of $P>0.5$. The last layer of $\mathcal{G}_{\boldsymbol{\Phi}}$ was a convolution layer with 3 kernels that yielded the transformation $\boldsymbol{\mathcal{T}}$ in x, y and z axes. To apply the estimated $\boldsymbol{\mathcal{T}}$ for deforming images and probabilistic segmentations, we adopted the spatial transformation function used in \cite{balakrishnan2019voxelmorph}. 

\subsubsection{Application to DWI.}
\label{ss:2.3}
Different DWI-derived metrics were used for the segmentation and registration component of the network (Fig.~\ref{fig1}), i.e., the diffusion tensor image (with six components) and fractional anisotropy (FA) image. For other applications, such as structure segmentation based on T1-weighted MRI, $\mathcal{F}_{\boldsymbol{\Theta}}$ and $\mathcal{G}_{\boldsymbol{\Phi}}$ could take the same inputs directly.

\section{Experiments and results}
\label{s:3}
\subsection{Material and preprocessing}
\label{ss:3.1}
 
\subsubsection{Material.} \label{sss:3.1.1}
The Rotterdam Study (RS) is a prospective and population-based study targeting causes and consequences of age-related diseases \cite{hofman2015rotterdam}. For this study, we used 8045 longitudinal DWI scans of 3249 individuals. The majority of these scans were repeatedly acquired in a time interval of 1--5 years (7770 scans of 3166 individuals), in which changes in brain microstructure are expected owing to aging. These long time-interval scans were matched into 6043 pairs by grouping any two time-points of the same individual. We used 5175 pairs for training ($D_{train}$), 200 pairs for validation ($D_{vali}$), and an independent cohort of 668 pairs for testing ($D_{test}$). The remaining scans were from 112 individuals who were scanned twice within a month, in which no changes in microstructure are expected. After exclusion of 15 individuals who had other visits in $D_{train}$, we used 97 pairs of short time-interval scans from 97 individuals for a reproducibility dataset ($D_{repro}$).

\subsubsection{MRI acquisition.} \label{sss:3.1.2}
Scans were acquired on a 1.5T MRI scanner (GE Signa Excite). DWI was scanned with the following parameters: $TR/TE = 8575 ms/82.6 ms$; imaging matrix of $64\times96$ (zero-padded in k-space to $256\times256$) in a field of view of $210\times210 mm^2$; 25 diffusion weighted volumes with a b-value of $1000 s/mm^2$ and 3 non-weighted volumes. The voxel size was resampled from $3.3\times2.2\times3.5 mm^3$ to $1 mm^3$, resulting in an image of $211\times210\times123$ voxels.

\subsubsection{Image preprocessing.} \label{sss:3.1.3}

DWI preprocessing \cite{li2018reproducible} included motion and eddy currents correction, diffusion tensors estimation using Levenberg–Marquard optimization, and computation of diffusion tensor imaging (DTI) measures such as FA. For each tract, an ROI was defined by taking the maximum bounding box based on the reference segmentation (section below). For the experiments, the forceps minor (FMI) tract is evaluated because it is functionally significant, i.e., related to aging and dementia, and not easy to be segmented. The ROI size for the FMI tract was $96\times64\times64$ voxels.

\subsubsection{Reference segmentations.} \label{sss:3.1.4}
 
The segmentation labels for model training and evaluation were generated using a published method \cite{de2015tract} consisting of probabilistic tractography and atlas information. The resulting tract-density images for each tract were normalized by division with the total number of tracts in the tractography run. Finally, tract-specific thresholds for the normalized density images were established by maximizing the reproducibility of FA measures. In the remainder of the paper, we denote the reference segmentations as $S^R$. 

\subsection{Experiments and evaluation metrics}\label{ss:3.2}
\subsubsection{Multistage pipelines.} \label{sss:3.2.1} 
We compared the proposed hybrid method ($*^H$) with two multistage pipelines ($*^{R,E}$ and $*^{N,E}$). First, $*^{R,E}$ denotes the reference segmentation method ($S^R$) in combination with a conventional registration algorithm using Elastix software ($\boldsymbol{\mathcal{T}}^E$) \cite{klein2010elastix}. Second, $*^{N,E}$ denotes a recent CNN-based segmentation method, Neuro4Neuro ($S^N$) \cite{li2018reproducible}, combined with Elastix. In Elastix (version 4.8), we used the B-spline based non-linear deformation, mutual information as similarity metric and a stochastic gradient descent optimizer with adaptive step size estimation. The paired samples t-test was used to test the statistical significance of below metrics.

Segmentation accuracy was quantified using the Dice coefficient (DC). For Neuro4Neuro, $accuracy^N = DC\big(S^R,S^N\big)$. For the hybrid method:
\begin{gather}
accuracy^H = DC\big(S^R, \mathcal{F}^H_{\boldsymbol{\Theta}}(I)\big)\label{eq4}.
\end{gather}

For each image pair, the consistency of segmentations was computed using the DC in two directions, i.e., both transforming the baseline to follow-up and transforming the follow-up to the baseline. For multistage pipelines, the transformation $\boldsymbol{\mathcal{T}}$ was estimated by aligning FA images using Elastix, e.g., $consistency^{N,E} = DC\big(S^N_t, S^N_s(\boldsymbol{\mathcal{T}}^E)\big)$. For the hybrid method, we tested the dataset bidirectionally since the segmentation in native space and that transformed from another time-point were available: 
\begin{gather}
consistency^H = DC\bigg(\mathcal{F}^H_{\boldsymbol{\Theta}}(I_t), \mathcal{F}^H_{\boldsymbol{\Theta}}(I_s)\big(\boldsymbol{\mathcal{T}}^H\big)\bigg).\label{eq9}
\end{gather} 

Using the dataset $D_{repro}$, we evaluated the scan-rescan reproducibility on segmentations and on tract-specific FA measures. The agreement of segmentations was quantified using the Cohen's kappa coefficient ($\kappa$):
\begin{gather}
\kappa^{H} = \kappa\bigg(\mathcal{F}^H_{\boldsymbol{\Theta}}(I_t), \mathcal{F}^H_{\boldsymbol{\Theta}}(I_s)\big(\boldsymbol{\mathcal{T}}^H\big)\bigg).\label{eq10}
\end{gather} 
Median of FA measures was computed for voxels inside the tract segmentation and used as the tract-specific measure. For each individual ($i$), the scan-rescan reproducibility of FA measures (FA$_{i1}$,FA$_{i2}$) was quantified by computing the error ($\epsilon$, Eq.~\ref{eq7}). A lower error indicates a higher reproducibility.
\begin{equation}\label{eq7}
\epsilon = \frac{1}{\frac{1}{2}} \frac{ |FA_{i1}- FA_{i2}|}{|FA_{i1}+ FA_{i2}|} 100 \%. 
\end{equation} 

\subsubsection{Registration and segmentation components.} \label{sss:3.2.2}
To assess the added value of the hybrid approach for registration and segmentation components, we designed separate step-wise experiments.

First, we assessed whether registration improves by joint optimization. We built a pure registration CNN (RegNet) using the same architecture as $\mathcal{G}_{\boldsymbol{\Phi}}$, optimized it with only $\mathcal{L}_{reg}$ and $\mathcal{L}_{def}$ terms \cite{balakrishnan2019voxelmorph}. Registration performance was compared between RegNet (independent optimization) and the registration component of hybrid method (joint optimization), and evaluated by registering the same segmentation image $S^R$, i.e., $S^R$+RegNet vs $S^R$+Hybrid. Registration accuracy was quantified by the spatial correspondence of the registered segmentations, i.e., DC($S^R_t, S^R_s(\boldsymbol{\mathcal{T}})$).

Second, we assessed whether segmentation improves by joint optimization. We built a pure segmentation CNN (SegNet) using the same architecture as $\mathcal{F}_{\boldsymbol{\Theta}}$, optimized it with only $\mathcal{L}_{seg}$ term. Segmentation performance was compared between SegNet (independent optimization) and the segmentation component of hybrid method (joint optimization), i.e., SegNet+Elastix vs Hybrid+Elastix. Segmentation accuracy was measured using the DC with reference segmentation ($S^R$). The consistency was evaluated by the spatial correspondence of the segmentations registered using Elastix, i.e., DC($S_t, S_s(\boldsymbol{\mathcal{T}}^E)$).

\subsubsection{Implementation.} \label{sss:3.2.3}
The experiments of model training and evaluation were performed on an NVIDA 1080Ti GPU and an AMD 1920X CPU. CNN-based methods were implemented using Keras-2.2.0 with a Tensorflow-1.4.0 backend. Models were trained using the Adam optimizer with an initial learning rate of $1e^{-4}$.

\subsection{Results} \label{ss:3.4} 

\begin{figure}[!]
\centering
\includegraphics[width=0.7\textwidth]{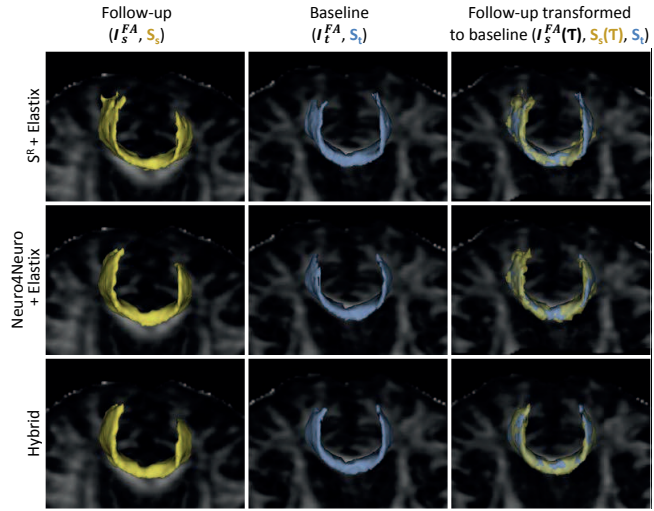}
\setlength{\belowcaptionskip}{-12pt}
\caption{Test results of the segmentation and registration. Colored structures indicate segmentations on the baseline (at 70 years old) and follow-up (26 months later) scans. The segmentation and FA image of the follow-up were transformed to the baseline. }\label{fig2}
\end{figure}

The segmentation accuracy, consistency and reproducibility of the proposed hybrid method were significantly higher than those of the multistage pipelines (Table~\ref{tab1}). The reproducibility of FA measures ($\epsilon=2\%$) was higher than that reported in the literature \cite{yendiki2016joint}  ($\epsilon=11\%$), and higher than their tailored longitudinal version ($\epsilon=5\%$), in which tracts were jointly reconstructed from both test-retest images. Figure~\ref{fig2} shows example results of scans with an average performance of the proposed method. Although all three methods showed reasonable results, the overlap of the transformed and target segmentations ($S_s(\boldsymbol{\mathcal{T}})$, $S_t$) was visually more consistent for the hybrid method than for the multistage pipelines.

\begin{table}[!]
\centering
\caption{Comparisons with the multistage pipelines. $S^R$, reference segmentation. $\kappa$, Cohen's kappa coefficient. $\epsilon$, scan-rescan error in FA measures. \textbf{Bold font} indicates a statistically significant improvement over the multistage pipelines ($p<0.01$).} \label{tab1}
\begin{tabular}{c|c|c|c|c}
\hline
Method & Accuracy$_{seg}$ &Consistency$_{seg}$ & $\kappa$ &$\epsilon  (\%)$\\
\hline
$S^R$ before registration  & -&  $0.31 \pm 0.16$ &$0.39 \pm 0.18$ &-\\
\hline
$S^R$+Elastix & - &  $0.55 \pm 0.09$ &$0.64 \pm 0.08$&$2.7\%$\\
Neuro4Neuro+Elastix & $0.67 \pm 0.08$ & $0.70 \pm 0.06$ & $0.76 \pm 0.06$ & $2.5\%$\\
Hybrid & $\textbf{0.70} \pm 0.07$ & $\textbf{0.73} \pm 0.11$  & $\textbf{0.78} \pm 0.08$&$2.3\%$\\
\hline
\end{tabular}
\end{table}

\begin{table}[!]
\centering
\caption{The added value of the hybrid method. Values in red cell indicate the mean $\pm$ SD of registration accuracy (DC($S^R_t, S^R_s(\boldsymbol{\mathcal{T}})$)) for the registration component. Those in blue cell indicate segmentation accuracy and consistency (DC($S_t, S_s(\boldsymbol{\mathcal{T}}^E)$)) for the segmentation component. \textbf{Bold font} indicates a statistically significant improvement over the other method within the colored group ($p<0.01$).}\label{tab2}
\begin{tabular}{c|c|c|c}
\hline
Method & Accuracy$_{seg}$ & Consistency$_{seg}$ & Accuracy$_{reg}$\\
\hline
$S^R$ + RegNet& - & - &\cellcolor[rgb]{1,0.85,0.85}$0.53 \pm 0.10$\\
$S^R$ + Hybrid & - & - &\cellcolor[rgb]{1,0.85,0.85}$\textbf{0.55} \pm \textbf{0.11}$\\
\hline
SegNet + Elastix & \cellcolor[rgb]{0.85,0.85,1}$0.69 \pm 0.08$ &\cellcolor[rgb]{0.85,0.85,1}$0.72 \pm 0.06$ & -\\
Hybrid + Elastix & \cellcolor[rgb]{0.85,0.85,1}$\textbf{0.70} \pm \textbf{0.07}$ &\cellcolor[rgb]{0.85,0.85,1}$0.72 \pm 0.06$ & -\\
\hline
\end{tabular}
\end{table}

Table~\ref{tab2} shows that both the registration and segmentation components of the hybrid CNN benefit from the joint optimization. For the registration task (red), the jointly optimized registration component of the hybrid method yielded a  significantly higher accuracy than the independently optimized RegNet on registering $S^R$. Accordingly, for the segmentation task (blue), the jointly optimized segmentation component of the hybrid method yielded a significantly higher segmentation accuracy than the independently optimized SegNet.  

\section{Discussion and conclusion}
\label{s:4}

We propose a novel hybrid deep learning framework for integrated segmentation and deformable registration in a single fast procedure. The framework was evaluated on longitudinal white matter tracts analysis using a large-scale diffusion MRI dataset. We show that the hybrid method leads to significantly higher accuracy, consistency and reproducibility of segmentation than multistage pipelines, and was orders of magnitude faster. Also, concurrent segmentation of structures and spatial alignment of time-points enables direct and consistent quantification of brain changes. Therefore, we expect the proposed method can open a novel way to support clinical and epidemiological analyses on understanding brain imaging changes over time. 

%
%
%
%
\bibliographystyle{splncs04}
\bibliography{Hybrid_model.bib}
\end{document}